\documentclass[iop,revtex4]{emulateapj}

\usepackage{natbib}
\usepackage{longtable}
\usepackage{rotating}
\usepackage[usenames,dvipsnames]{color}
\usepackage{ulem}

\def\arcsec{\hbox{$^{\prime\prime}$}}

\def\cm2{cm$^{-2}$}

\def\nh3{NH$_3$}
\def\n2h{N$_2$H$^+$}

\def\13co{$^{13}$CO}
\def\c18o{C$^{18}$O}
\def\hc3n{HC$_3$N}
\def\h2{H$_2$}
\def\nh{n(H$_2$)}

\newcommand{\msun}{\ensuremath{M_{\odot}}}
\newcommand{\mbh}{\ensuremath{M_{\rm BH}}}

\newcommand{\ha}{H\ensuremath{\alpha}}

\newcommand{\etal}{et~al.}

\newcommand{\lbol}{\ensuremath{L{_{\rm bol}}}}
\newcommand{\ledd}{\ensuremath{L{_{\rm Edd}}}}
\newcommand{\lratio}{\ensuremath{\ell}} 

\begin{document}

\slugcomment{}

\shorttitle{Low-Mass AGN Fundamental Plane} \shortauthors{Qian, et al.}

\title{Low-mass Active Galactic Nuclei on the Fundamental Plane of Black Hole Activity}


\author{Lei Qian\altaffilmark{1}, Xiao-Bo Dong\altaffilmark{2},
Fu-Guo Xie\altaffilmark{3},
Wenjuan Liu\altaffilmark{2}, and Di Li\altaffilmark{1} }

\affil{}
\altaffiltext{1} {CAS Key Laboratory of FAST, NAOC, Chinese Academy of Sciences, Beijing 100012, China;\  \mbox{lqian@nao.cas.cn}}

\altaffiltext{2}{Yunnan Observatories, Chinese Academy of Sciences, Kunming, Yunnan 650011, China;
Key Laboratory for the Structure and Evolution of Celestial Objects, Chinese Academy of Sciences,
Kunming, Yunnan, 650011, China;\  \mbox{xbdong@ynao.ac.cn} }

\altaffiltext{3}{Key Laboratory for Research in Galaxies and Cosmology, Shanghai Astronomical Observatory,
Chinese Academy of Sciences, 80 Nandan Road, Shanghai 200030, China;\  \mbox{fgxie@shao.ac.cn} }

\begin{abstract}
It is widely known that in active galactic nuclei (AGNs) and black hole X-ray binaries (BHXBs),
there is a tight correlation among their radio luminosity ($L_R$), X-ray luminosity ($L_X$) and BH mass ($\mbh$),
the so-called `fundamental plane' (FP) of BH activity.
Yet the supporting data are very limited in the $\mbh$ regime between stellar mass (i.e., BHXBs)
and 10$^{6.5}$\,\msun\ (namely, the lower bound of supermassive BHs in common AGNs).
In this work, we developed a new method to measure the 1.4 GHz flux
directly from the images of the VLA FIRST survey,
and apply it to
the type-1 low-mass AGNs in the \cite{2012ApJ...755..167D} sample.
As a result, we obtained 19 new low-mass AGNs  for FP research
with both \mbh\ estimates ($\mbh \approx 10^{5.5-6.5}$\,\msun), reliable X-ray measurements,
and (candidate) radio detections,
tripling the number of such candidate sources in the literature.
Most (if not all) of the low-mass AGNs follow the standard radio/X-ray correlation
 and the universal FP relation
fitted with the combined dataset of BHXBs and supermassive AGNs by \citet{2009ApJ...706..404G};
the consistency in the radio/X-ray correlation slope
among those accretion systems
supports the picture 
that the accretion and ejection (jet) processes
are quite similar in all accretion systems of different \mbh.
In view of the FP relation, we speculate that  
the radio loudness $\mathcal{R}$ (i.e., the luminosity ratio of the jet to the accretion disk) of AGNs
depends not only on Eddington ratio, but probably also on \mbh. 
\end{abstract}

\keywords{
galaxies: active ---
galaxies: intermediate-mass black holes ---
galaxies: jets --- radio continuum: galaxies --- X-rays: galaxies ---
black hole physics
}

\section{Introduction}

Highly collimated relativistic jets, most evident in radio emission,
are a remarkable observational phenomenon of active galactic nuclei (AGNs).
Consensus has been reached that jets are a direct consequence of the accretion process onto black holes (BHs).
Decades after their discovery, however,
it is still a fundamental question in accretion theory
regarding the mechanism of the launching, accelerating, collimating and propagating of a jet
(for reviews see, e.g., \citealt{1998ARA&A..36..539F,2010LNP...794..233S,2015SSRv..191..441H}).
It also remains open as to what physical factors govern the jet power and thus the radiative output.
The accretion rate is obviously a key factor, but additional factors should also play their roles (see below).

Observationally, in the studies of  the radio emission of AGNs,
radio loudness $\mathcal{R}$ is widely adopted to characterize
the relative radiative importance of the jet to the accretion disk (AD).
It is defined as the ratio of the radio luminosity (traditionally at \mbox{5\,GHz})
to either the UV luminosity (monochromatic,  at the $B$ band)
or the X-ray luminosity (integrated in the 2--10 keV range);
i.e., $\mathcal{R}=L_\nu(\mbox{\rm 5\,GHz})/L_\nu(B)$ \citep{1989AJ.....98.1195K} or
$\nu L_\nu(\mbox{\rm 5\,GHz})/L_X(\mbox{\rm 2--10\,keV})$ \citep{2003ApJ...583..145T}.
There have been a lot of observational investigations on the relation between $\mathcal{R}$
and other AGN parameters such as
AGN luminosity or the Eddington ratio \citep[$\lratio \equiv \lbol/\ledd$; see][]{2002ApJ...564..120H}
 \footnote{\,The parameter Eddington ratio (\lratio) is defined as the ratio
 between the bolometric luminosity (\lbol) and
the Eddington luminosity ($L_{\rm Edd} = 1.3\times 10^{38}\, (M_{\rm BH}/M_{\odot})\,{\rm erg\,s^{-1}}$).
In accretion-powered radiation systems, \lratio\ is often
referred to as the dimensionless accretion rate $\dot{m}$
(the mass accretion rate normalized by the Eddington accretion rate,
$\dot{m} \equiv \dot{M}/\dot{M}_{\rm Edd}$),
as $\dot{m}$ is not a direct observable.  Yet, the two notations are different,
both in meaning and in scope of application; see Footnote~8 of \cite{2011ApJ...736...86D}. },
BH mass \citep[\mbh; see][]{2000ApJ...543L.111L, 2002ApJ...564..120H},
host-galaxy morphology \citep[as a proxy of BH spin; see][]{2007ApJ...658..815S},
galactic environment \citep[e.g.,][]{2009ApJ...697.1656S}, etc. 
Unfortunately, no consensus has been reached
\citep[see the references above;
also][]{2008ApJ...685..801Y,2016A&ARv..24...10T,2016ApJ...833...30C,2017MNRAS.466..921C}.
For instance,
is the apparent dichotomy between radio quiet ($\mathcal{R} < 10$) and radio loud intrinsic or not?
What is the primary driver of $\mathcal{R}$, \mbh\ or \lratio\ (or both)?

On the other hand,
empirical relations have been explored in the line of the coupling between accretion disk and jet.
\citet{2003MNRAS.345.1057M} extended the work of the radio/X-ray correlation
discovered in BH X-ray binaries (BHXBs) (\citealt{2003A&A...400.1007C,2013MNRAS.428.2500C}) to AGNs,
and discovered a linear (in logarithmic space) correlation
among $L_R$, $L_X$ and $\mbh$,
which is usually called the `fundamental plane' (FP) of BH activity
(see also \citealt{2004A&A...414..895F,2008ApJ...688..826L,2009ApJ...706..404G,2014ApJ...788L..22G,
2015MNRAS.453.3447D,2015MNRAS.447.1289P,2016ApJ...818..185F,2017ApJ...836..104X}).
Note that in FP studies only continuous/steady jets rather than transient/episodic jets are considered
(see \citealt{2009MNRAS.396.1370F} for the classification of these two types of jets in X-ray binaries).

The FP can be written as
\begin{equation}
\log L_{R}=\xi_X \log L_{X}+\xi_M \log M_{\rm BH} +C ~ .
\label{FP}
\end{equation}
The best-fit parameters achieved by \citet{2003MNRAS.345.1057M} are
$\xi_X=0.60^{+0.11}_{-0.11}$, $\xi_M=0.78^{+0.11}_{-0.09}$, and $C=7.33^{+4.05}_{-4.07}$.
Later works found that the FP is remarkably tight;  
and individual systems reveal that the FP or the standard radio/X-ray correlation holds
for a large dynamic range in $L_{\rm X}/L_{\rm Edd}$,
even for the so-called quiescent BHXBs (\citealt{2014MNRAS.445..290G,2017ApJ...834..104P}).
They are consistent with the theoretical expectation of the coupled accretion--jet models
(\citealt{2005ApJ...629..408Y}; see also \citealt{2003MNRAS.343L..59H}).
Furthermore, FPs with different coefficients
seem to apply in bright, radiatively efficient AGNs 
(see, e.g., \citealt{2014ApJ...787L..20D}),
as well as in faint, jet-emission-dominated AGNs
(see, e.g., \citealt{2009ApJ...703.1034Y,2017ApJ...836..104X}).
However,
several BHXBs (\citealt{2011MNRAS.414..677C,2013MNRAS.428.2500C})
and one AGN, NGC 7213 (\citealt{2011MNRAS.411..402B,2016MNRAS.463.2287X}),
may be yet classified as ``outliers" to the standard FP,
as their individual variability exhibits hybrid radio/X-ray correlations
rather than a single power-law relation with a constant index $\xi_X$ (cf. \S3.1).
Following the discovery of the FP,  
there have been intense discussions in the literature,
particularly focusing on understanding the physics of the AD--jet
coupling in AGNs (of different accretion rates) with the insights
from the accretion states and state transitions of BHXBs
(e.g., \citealt{2006MNRAS.372.1366K}), and on applying the FP to estimate
\mbh\  
(e.g., \citealt{2009ApJ...706..404G}).
Apart from those implications and applications,
even if the FP were valid only statistically (namely in an ensemble sense)  for AGNs,
it provides a coherent interpretation to
the aforementioned observed phenomena concerning $\mathcal{R}$,
at least as an empirical induction that may instill a deeper insight.

In the FP studies, there is a clear gap in \mbh\ between the BHXBs with stellar-mass BHs
and common AGNs with supermassive BHs (SMBHs, $\mbh \gtrsim 10^{6.5} \msun$).
As noted in  \citet{2014ApJ...788L..22G},
accreting systems with $\mbh \approx 10^{2-6}$\,\msun\
(a range that includes low-mass AGNs) are crucial.
There are several reasons for this.
First, these systems will fill the mass gap of current FP research.
Second, the evolutionary timescale of accretion onto the BHs is likely systematically shorter
than that in common AGNs.
Consequently, 
it is possible in the future to investigate the FP in \textit{individual} low-mass AGNs of very small \mbh,
i.e., the radio/X-ray correlation at given \mbh\ values exhibited in the variability of individual sources,
similar to the case of BHXBs;  
see, e.g., NGC\,7213 (Bell \etal\ 2011), and NGC\,4395 (King \etal\ 2013).
The radio/X-ray correlation of individual sources will provide a cleaner environment,
where systematic uncertainties caused by the BH spin and relativistic beaming effect due to different viewing angles are eliminated.
\citet{2014ApJ...788L..22G} for the first time used low-mass AGNs to check
the two FP relations of \citet{2009ApJ...706..404G},
one based on a SMBH-only sample and the other based on a combined SMBH$+$BHXB sample.
Their result suggested that there exists a universal FP
that is valid for accreting systems with \mbh\ of all scales , i.e., SMBHs, low-mass AGNs and stellar-mass BHs.
However, the low-mass AGN sample size of \citet{2014ApJ...788L..22G}
is rather limited, with only 10 data points (including 3 sources with only upper limits on radio or X-ray measurements).
Further investigations of the FP with more low-mass AGNs are needed.

In this work, we developed a new method to obtain the radio measurement directly from the images of
the VLA FIRST survey (\citealt{1995ApJ...450..559B}), and applied it to the low-mass AGN sample of \cite{2012ApJ...755..167D}.
This results in 19 new low-mass AGNs for FP research,
with known virial \mbh\ estimated from their broad emission lines,
firm X-ray detection, and radio detection with S/N $>3$ (8 of the 19 objects can be regarded as reliable radio sources).
The method and our measurements, as well as the collection of low-mass AGNs in the literature suitable for FP research,
are presented in \S2.
\S3 presents our data analysis and results.
\S4 gives a brief summary and plans for the future works.
Throughout the paper, we assume a flat cosmology with
$H_0=70\ \rm km\ s^{-1}\ Mpc^{-1}$, $\Omega_{\rm m}=0.3$, and $\Omega_{\lambda}=0.7$.

\section{Data of low-mass AGNs}

The low-mass AGN sample used in this work includes two datasets.
The main dataset is our measurements based on VLA FIRST images and archival X-ray observations,
comprising 19 sources with known virial mass \mbh.
With measurements of all the three parameters ($L_R$, $L_X$ and \mbh),
they  represent significant increment to
the research of BH FP.
This new dataset is listed in Table~1.

For completeness,
we also collect all the low-mass AGNs in the literature with reliable measurements of  $L_R$ and $L_X$ and
having \mbh\ estimates that are not from the BH FP relation.
This supplementary literature dataset includes 10 sources in total (see Table~2).

The entire low-mass AGN sample includes 29 sources, 18 of which have well-measured radio luminosities.
The sources of the entire sample are moderately bright, as shown in Figure~\ref{rxedd} (the $x$-axis).
With four exceptions, all sources have $10^{-3}\la L_X/\ledd \la 1$,
clustering in the range $L_X/\ledd \approx 10^{-2}$\,--\,$10^{-0.5}$.
According to the $L_X/\ledd \approx 1\times 10^{-3}$ separation between bright AGNs and low-luminosity AGNs
based on their distinctive properties (e.g. \citealt{2008ARA&A..46..475H,2015MNRAS.447.1692Y}),
the low-mass AGNs here mainly belong to the bright AGN category.

Below we describe the radio and X-ray measurements of the new dataset, and the compilation of the literature dataset.

\subsection{Our measurements}
\label{sample}

Our parent low-mass AGN sample is the 309 broad-line AGNs with $\mbh < 2 \times 10^6 \msun$
compiled by \citet{2012ApJ...755..167D}.
The $\mbh$ was derived from the luminosity and width of the broad H${\alpha}$ emission line,
using the virial formalism calibrated by \cite{2007ApJ...670...92G} based on single-epoch spectra.
The statistical uncertainty of the estimated $\mbh$ should be around 0.3\ dex typically
(see, e.g., \citealt{2009ApJ...707.1334W} for the uncertainty estimation for an AGN sample);
yet the uncertainty for some individual sources can be as large as 1 order of magnitude
(e.g., Vestergaard \& Peterson 2006).
We set the uncertainty of \mbh\ (and accordingly the Eddington luminosity \ledd)  to be 0.6~dex
in the subsequent fitting (\S3.1) and plottings (\S3.1 and \S3.2).  

Among the sources of \citet{2012ApJ...755..167D}, 288 sources are covered by the VLA FIRST survey.
For these sources, we fit the FIRST images,
\footnote{https://third.ucllnl.org/cgi-bin/firstcutout}
and measure the fluxes and the corresponding rms (root of mean square) noises
(the details of the fitting and noise determination can be found in \S\ref{sec:rflux}).

The sources with flux greater than 3 times the rms noise (namely S/N $>3$)
are deemed to have radio detections.
This criterion is a tradeoff between minimizing false detections
and maximizing the number of reliable radio sources
(or candidates of high probability).
We will analyze this criterion at the end of this subsection and check it in \S3.
There are 52 such radio-detected sources.
Then we match them to the X-ray archive, NASA's HEASARC,
\footnote{http://heasarc.gsfc.nasa.gov}
and find 22 sources that we can obtain their X-ray fluxes (\S2.1.2).

Among the 22 low-mass AGNs with both radio and X-ray fluxes,
there are 3 sources that have already been included in the literature
(e.g., included in the low-mass AGN dataset used by \citealt{2014ApJ...788L..22G};
see also Table~2 of \citealt{2012ApJ...753..103N});
they are J0914$+$0853, J1240$-$0029 (namely GH10 after \citealt{2004ApJ...610..722G}),
and J1324$+$0446.
Excluding those 3 sources (they are listed in the literature dataset instead; see Table~2),
our new dataset includes 19 sources.
\newline

Here we must evaluate our criterion of radio detection, S/N $>3$.
We set such a criterion instead of the commonly used  flux limit S/N $>5$
(or called 5$\sigma$  if the noise is random and Gaussian),
out of the tradeoff between reliability and the purpose
to select as many as possible radio-detected sources or candidates.
Radio-detected low-mass AGNs are rather rare,
and thus even the selection of candidate radio sources has its own merit.
Assuming the noise of the FIRST images is Gaussian,
the trial penalty, namely the probability of
mistaking one or more random fluctuations as radio source(s) with S/N $>3$
out of the parent sample of 288 objects covered by the FIRST survey,
is  $1-\left(\int^{3\sigma}_{-\infty} G(x)\, \mathrm{d} x\right)^{288} = 0.32$.
Here $G(x)$ is the Gaussian probability density function,
zero-centered and with a standard deviation $\sigma$.
A chance probability of 0.32 is fairly large.
To be worse, there are often correlated errors in radio images,
and thus the noise is not purely random Gaussian and
the false-detection probability would be greater than the above estimated trail penalty;
this is the very reason why the conservative flux limit of $5\sigma$ is commonly used.
Certainly, on the other hand, our estimation of noise (namely rms; see \S2.1.1)
is not merely the random component, but is able to incorporate other error sources to some degree.

In order to make up the shortcoming of the S/N $>3$ criterion,
we divide the 19 sources into two groups:
8 sources with S/N $>4.43$ (including 4 sources with S/N $>5$) are grouped as the reliable radio detections,
and the remaining 11 with radio $3<$ S/N $<4.43$,
conservatively speaking, are only candidates.
The dividing S/N of 4.43 is set in terms of trial penalty, as follows. 
For one source, assuming random Gaussian noise
the chance probability of false detection associated with the $S/N >3$ criterion
is \mbox{$1-\int^{3\sigma}_{-\infty} G(x)\,\mathrm{d} x$} $= 0.0013$.
We now require the chance probability for our parent sample of 288 objects to be the same level, i.e.,
\mbox{$1-\left(\int^{n\sigma}_{-\infty} G(x) \, \mathrm{d} x\right)^{288}$} $\approx 0.0013$,
then we get $n = 4.43$.
In the subsequent analyses (\S3.1 and \S3.2), we will compare the candidate radio sources
with our reliable sources and the literature sources, in the radio/X-ray correlation
and in the FP. We find no difference between the two groups of sources.

\subsubsection{Radio Flux}
\label{sec:rflux}

We adopt a new method to obtain the (faint) radio flux for as many low-mass AGNs as possible.
This method was first used by \cite{2017ApJ...837..109L}, to measure the 1.4\,GHz flux
directly from the VLA FIRST image, when the radio emission is faint and below the flux threshold (1 mJy)
set to the official FIRST catalog (\citealt{1995ApJ...450..559B,1997ApJ...475..479W}).
Low-mass AGNs are generically radio quiet ($\mathcal{R} < 10$), with only a few ($<6$\%) being radio loud
(\citealt{2006ApJ...636...56G}); this is supposedly due to their relatively high accretion rate by selection
(as well as small \mbh\ compared with common AGNs with SMBHs; cf. \S4 below).
Among the 288 low-mass AGNs covered by the FIRST survey, only 17 are included in the FIRST catalog.

We fit a two-dimensional Gaussian to the FIRST images of every sources.
The potential radio sources are assumed to be point-like, with the Gaussian FWHM set to
be the beam size (5\farcs4).
The center of the Gaussian is fixed to be the optical position of the broad-line nucleus determined by
the Sloan Digital Sky Survey (SDSS; \citealt{2000AJ....120.1579Y}).
The only free parameter is the flux of the point source.

In addition, a CLEAN bias always makes an underestimate of the flux,
which is typically \mbox{0.25\,mJy} for point sources
(see \S4.3 of \citealt{1997ApJ...475..479W} and \S7.2 of \citealt{1995ApJ...450..559B}).
We thus correct this bias for the best-fit flux by adding 0.25 mJy, as
all versions of the official FIRST catalog released after 1995 October did (\citealt{1997ApJ...475..479W}).
\footnote{See also the handling of the CLEAN bias in the current version of the FIRST catalog
(dated 2014 December 17, which is used in our present work):\\
http://sundog.stsci.edu/first/catalogs/readme\_14mar04.html\#cleanbias\,. We caution that
such a correction of 0.25 mJy might be inappropriate when the radio sources are
close to the rms level; this should be tested by future deeper observations.}
As the CLEAN bias arises from that
CLEAN algorithm steals flux from discrete sources and spreads it around the image,
it also has some influence on the rms noise, which is not yet well understood.
We simply  measure the rms noise in an empty region of size $9\arcsec\times 9\arcsec$,
20\arcsec\ away from the center of the Gaussian,
and take it as the uncertainty of the \mbox{1.4\,GHz} flux.

Note that our thus-measured rms values are the actual ones directly from the final co-added images,
different from what were used in the FIRST catalog construction (namely the so-called 5$\sigma$ flux threshold)
and listed in the catalog (the `RMS' column).
The latter was based on the weighted combination of noise values derived from
the whole-image rms for each grid map that contributes to that image,
as displayed in the rms sensitivity map of the FIRST coverage; \cite{1997ApJ...475..479W}
noted that ``it [the coverage-map rms value] should not be used
to establish a definitive upper limit to the radio flux density from a given location in the sky;
rather, the flux density in the relevant co-added image should be measured directly.''

We consider the sources with the flux (prior to the correction for the CLEAN bias) higher than 3 times the rms noise (namely S/N $>3$)
as candidate radio detections.
Note that in the S/N calculation the flux is the one prior to the correction for the CLEAN bias. 
There are 52 such radio-detected sources (including the 17 already in the FIRST catalog),
22 of which have archival X-ray data.
We derived their radio flux at 5 GHz from our measured 1.4 GHz flux.
Since our sources are relatively bright (namely actively accreting) with X-ray Eddington ratio $10^{-3} < L_X/\ledd < 1$,
we adopt the typical spectral index of bright AGNs,
$\alpha_R = -0.5$ (defined as $F_\nu \propto \nu^{\alpha_R}$), for the conversion.
We simply assume a 20\% uncertainty in $\alpha_R$,
which would cause a 13\% uncertainty in the 5 GHz flux.
\newline

Admittedly the validity of this new method needs to be tested.
For this purpose, we collect radio point sources that have radio flux data
and are covered by the FIRST survey,
use our new method to measure their radio fluxes, and then make the comparison.
We limit our method and the comparison for radio point-like sources only,
to minimize the contamination
of the radio emission from the host galaxies.
The point-like sources are selected in two ways.
One part is the 6 unresolved sources among the aforementioned 17 low-mass AGNs included in the FIRST catalog;
they are selected to have the beam-corrected major-axis FWHM $<1\farcs5$
(the parameter `\textit{Deconv.\,MajAx}' in the catalog).
The other part is the 6 sources that are in the literature dataset (Table~2)
and are covered by the FIRST survey
(but excluding  those with  \textit{Deconv.\,MajAx} $\geqslant 1\farcs5$ in the FIRST catalog).
Most of the sources are faint, close to the 1 mJy threshold of the FIRST catalog.
\footnote{When \cite{2014ApJ...788L..22G} selected the targets for their \mbox{5\,GHz} observation,
they found that part of their 10 radio sources were not in the FIRST catalog at that time.
All the 10 radio sources now are included in the current FIRST catalog
(dated 2014 Dec 17; http://sundog.stsci.edu/cgi-bin/searchfirst\,).}
The comparison is summarized in Table~\ref{radio_flux}.

We can see that the values by our method ($f_{\rm our}$) are in good agreement (all within a factor of
$\leqslant$\,1.3)
with the fiducial values ($f_{\rm cat}$; from the FIRST catalog).
The mean and standard deviation of the relative difference $(f_{\rm our}-f_{\rm cat})/f_{\rm cat}$  are $-0.11$
and $0.10$, respectively.
That is, the systematic error and the random uncertainty of our measured flux are both on the level of 10\% only.
Therefore it is reliable to apply our method to point-like sources to obtain their radio fluxes.
The standard error of the estimated mean (namely, the systematic offset) is only 2.9\%,
meaning that the systematic offset is fairly stable. Thus we correct
this offset of $-11$\% from our measured fluxes in the subsequent fitting (\S3.1)
and plottings (Figures~1 and 2). 

\subsubsection{X-ray Flux}

The $3\sigma$ confidence interval of the positional uncertainty of FIRST is about 1\farcs8
\citep{1995ApJ...450..559B}.
The $3\sigma$ confidence intervals of the positional uncertainties
of XMM-Newton and Chandra are
4\arcsec\ \citep{2007MNRAS.382..279P} and 2\farcs7
\citep{2007ApJS..169..401K}, respectively.
The matching radii
of FIRST sources to XMM-Newton and Chandra sources
are set to the square root of
the quadratic sum of the $3\sigma$ confidence intervals of FIRST and respective X-ray positions,
i.e. 4\farcs4, and 3\farcs2, respectively.
As to X-ray sources detected by ROSAT,
because the positional uncertainty of ROSAT is fairly large, with 1$\sigma$ being 20\arcsec\
\citep{1993AdSpR..13..391V},
we simply set a conservative matching radius of 20\arcsec.
Among the 52 radio sources, 22 have X-ray  detections (including 3 sources in the literature dataset).
The largest offset between the matched ROSAT sources (totaling 8) and their FIRST counterparts is 12\arcsec,
which is large and liable to false matching.
Thus, if a source has observations of sufficient quality by multiple missions, we adopt the data
with the best spatial resolution (or equivalently, with the best positional accuracy),
namely in the descending order of Chandra, XMM-Newton and ROSAT.
Finally, of the 19 sources in our new dataset, 10 sources adopt Chandra data,
3 adopt XMM-Newton data, and the rest 6 adopt ROSAT data (see Table~1).
The final adopted X-ray sources turn out to have the offset distances to their optical positions
within 6\arcsec\ (ROSAT), 2\arcsec\ (XMM-Newton) and 2\arcsec\ (Chandra), respectively.
The small offsets of the ROSAT matches with respect to their matching radius
are owing to the small number (6 sources).
Such positional offsets are roughly within the optical extent of their SDSS images.
We visually inspect their various optical images available in the NED%
\footnote{http://ned.ipac.caltech.edu}, and find no ambiguous sources coinciding within their offset distances.

The X-ray flux (or count rate) and its uncertainty are retrieved from HEASARC.
The flux is measured in the energy range of 0.3 to 8 keV for Chandra, and in 0.2 to 12 keV for XMM-Newton;
the count rate of ROSAT is measured in the energy range of 0.1 to 2.4 keV.
With these data, we then use the WebPIMMS\footnote{heasarc.gsfc.nasa.gov/cgi-bin/Tools/w3pimms/w3pimms.pl} of HEASARC
to convert the X-ray flux or count rate to the flux in the energy range of 2--10 keV,
assuming an absorbed power-law form with photon index taken to its typical value of $2$.
We simply assume a 10\% uncertainty in photon index in the subsequent error analysis,
which would cause a 15\% to 30\% uncertainty in the 2--10 keV flux.
For the absorption, we only consider that from our Galaxy, and the Galactic hydrogen column density is obtained
with the $N_{\rm H}$ tool
\footnote{http://heasarc.gsfc.nasa.gov/cgi-bin/Tools/w3nh/w3nh.pl} of HEASARC with LAB map.
Two sources (J0824$+$3800 and J1347$+$4743) were also included in
\cite{2015ApJ...808..163P}; their measured fluxes based on the XMM-Newton data
agree well with our measurements.

\subsection{Sources from the literature}
\label{literature}
In Table~\ref{literaturetable}, we list all the low-mass AGNs for the FP studies in the literature.
These include the 7 sources of \citet{2014ApJ...788L..22G} that had firm detections in both radio and X-ray
(with the 3 upper-limit sources dropped) and had virial \mbh\ from \cite{2007ApJ...670...92G}.
In addition, we also include
NGC 4395 and NGC 404 \citep{2012ApJ...753..103N}, and Henize 2-10 \citep{2016ApJ...830L..35R}.
The \mbh\ of NGC 4395 is obtained by reverberation mapping,
and the \mbh\  of NGC 404 is obtained by dynamical measurements.
The \mbh\ of Henize 2-10 is estimated from the stellar mass of the host galaxy.
See the above references for the details of the radio and X-ray measurements and the \mbh\ estimation.

\section{Low-mass AGNs on the fundamental plane}

\subsection{The radio/X-ray correlation}

\begin{figure}[tbp]
\centering
\includegraphics[width=8.5cm]{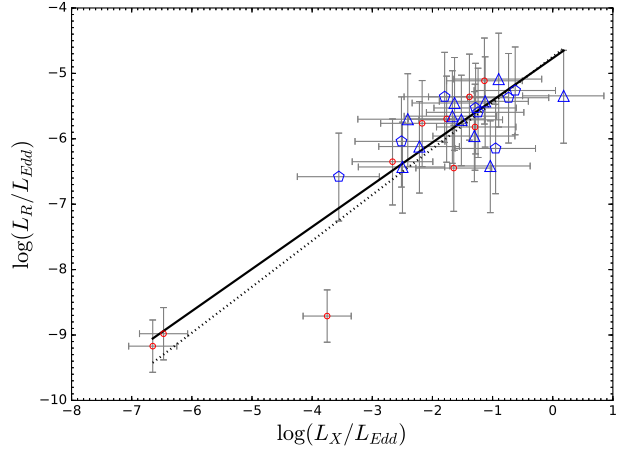}
\caption{Radio/X-ray correlation (in Eddington unit; $L_R/\ledd$---$L_X/\ledd$) of low-mass AGNs.
Our new sources are shown in two groups
as blue triangles (sources with flux in between $3\sigma$ and $4.43\sigma$) and
blue pentagons (sources with flux higher than $4.43\sigma$), respectively,
with $\pm1\sigma$ uncapped error bars;
the literature sources are denoted as red open circles, with $\pm1\sigma$ capped error bars.
Note that these three groups have similar distributions on this plot.
The dotted line [$\log (L_R/{\ledd}) = (0.70 \pm 0.05)\, \log (L_X/{\ledd}) -(4.75 \pm 0.18)$]
and solid lines [$\log (L_R/{\ledd}) = (0.64 \pm 0.04)\, \log (L_X/{\ledd}) -(4.77 \pm 0.11)$]
are the best fits with and without the `outlier' \mbox{NGC\,4395} (at $\log\,L_X/{\ledd} = -3.75$) included.
}
\label{rxedd}
\end{figure}

Before exploring the FP of BH activity, we first examine the radio/X-ray correlation among low-mass AGNs.
We consider their radio and X-ray luminosities in terms of Eddington unit, i.e. $L_R/\ledd$ and $L_X/\ledd$,
in order to reduce the impact of $\mbh$.
Figure~\ref{rxedd} shows the $L_R/\ledd$\,--\,$L_X/\ledd$ relationship,
where the sources of the new dataset are shown
as pentagons (the 8 reliable radio sources with S/N $>4.43$; see \S2.1)
or  triangles (the 11 candidate radio sources with $3<$ S/N $<4.43$),
while the literature dataset are shown as open circles.
We fit the entire sample with a single power-law model (i.e., linear in the log--log scale),
using the LINMIX\_ERR program (\citealt{2007ApJ...665.1489K}) that
accounts for measurement errors in both axes.
The systematic offset of our measured $L_R$ values with respect to the fiducial ones ($-11$\%; see \S2.1.1)
is corrected in the fitting.
The total uncertainty of $L_R$ is the quadrature sum of the following three terms:
the rms noise, the error introduced by the assumed $\alpha_R$, and the random uncertainty with respect to the fiducial
(see \S2.1.1 for the details of the three terms).
The total uncertainty of $L_X$ is the quadrature sum of the following two terms:
the uncertainty from the archive and the error introduced by the assume photon index (see \S2.1.2 for the details).
The uncertainty of \ledd\ is simply 0.6 dex (see \S2.1).
The fitting result is as follows (the dotted line in Figure~\ref{rxedd}),
\begin{equation}
\label{eq_LrLx0}
 \log (L_R/{\ledd}) = (0.70 \pm 0.05)\, \log (L_X/{\ledd}) -(4.75 \pm 0.18)  ~ ,
\end{equation}
with a reduced $\chi^2=0.56$. 
We can see from Figure~\ref{rxedd}  that a data point at $\log\,L_X/{\ledd} = -3.75$
deviates from the best-fit line
by about 3$\sigma$; this outlier is \mbox{NGC\,4395}.
When \mbox{NGC\,4395} is excluded, the best fit becomes
\begin{equation}
\label{eq_LrLx}
 \log (L_R/{\ledd}) = (0.64 \pm 0.04)\, \log (L_X/{\ledd}) -(4.77 \pm 0.11)  ~ ,
\end{equation}
with a reduced $\chi^2=0.25$;  
see the solid line in Figure~\ref{rxedd}.
Such a small reduced $\chi^2$ indicates that the uncertainties of the data are over-estimated to some degree. 
This best fit is close to the $L_R\propto L_X^{0.62}$ relation
reported in GX 339-4, a typical BHXB \citep{2013MNRAS.428.2500C},
implying that all the low-mass AGNs (probably except \mbox{NGC\,4395}) are standard ones
(namely, obeying a single power-law $L_R\propto L_X^{\xi_X}$) rather than ``outliers"
in terms of the radio/X-ray correlation of BH accreting systems (cf. \S1).


In order to test the difference between the candidate sources (3 $<$ S/N $<$ 4.43)
and the well-measured sources (S/N $>$ 4.43, excluding NGC 4395),
we exclude the candidate sources as well as  NGC 4395, and perform the fitting again.
The best fit is almost the same as Eq.~\ref{eq_LrLx}, being
$\log (L_R/{\ledd}) = (0.66 \pm 0.04)\, \log (L_X/{\ledd}) -(4.69 \pm 0.14)$.


We note in passing that the outlier \mbox{NGC\,4395} deserves further investigation in the future.
It has a reliable \mbh\ measurement by reverberation mapping method (\citealt{2005ApJ...632..799P}).
From a joint monitoring in radio (VLA) and X-ray ({\it Swift}/XRT) in 2011,
\mbox{NGC\,4395} seemed to follow a flat radio/X-ray correlation, i.e. $L_R\propto L_X^{\sim 0}$ (\citealt{2013ApJ...774L..25K}).
Although that result was not robust due to the very limited dynamic range in both $L_R$ and $L_X$,
it suggested that \mbox{NGC\,4395} might be a source that follows
the flat branch of the hybrid radio/X-ray correlation (\citealt{2016MNRAS.456.4377X};
cf. NGC 7213, \citealt{2016MNRAS.463.2287X}).

\subsection{The fundamental plane}

\begin{figure*}[tbp]
\centering
\includegraphics[width=18.0cm]{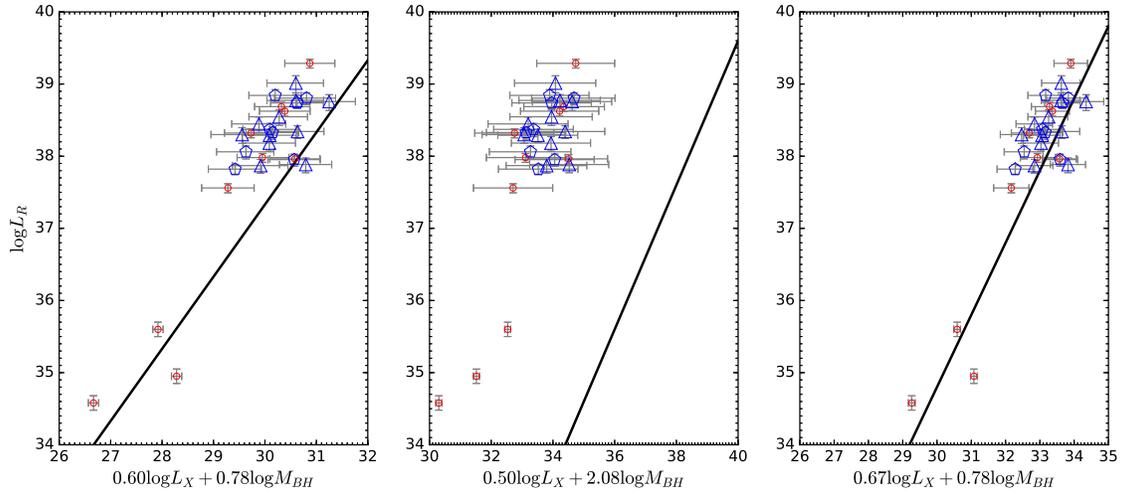}
\caption{Low-mass AGNs in the well-known fundamental planes of black hole activity.
The three fundamental planes (solid lines) from left to right are
the \citet{2003MNRAS.345.1057M} relation,
and the SMBH-only and the universal  relations of \cite{2009ApJ...706..404G}, respectively.
See the text in \S3.2 for the detail.
Symbols are the same as in Figure~\ref{rxedd}.
}
\label{fp_large_range}
\end{figure*}

We then examine the low-mass AGNs in the FP of BH activity.
Because our sample is still not large and the data (particularly the radio fluxes) demand to be refined,
in this work we refrain from fitting the data to get a new relation.
Instead, we take the same approach as \citet{2014ApJ...788L..22G},  
by examining our data with respect to several well-known FP relations in the literature.
Three FPs are considered: the \citet{2003MNRAS.345.1057M} relation ($\xi_X=0.60$, $\xi_M=0.78$, and $C=7.33$),
and the SMBH-only (i.e., fitted with SMBH systems only; $\xi_X=0.50$, $\xi_M=2.08$, and $C=0.40$) and the universal
(i.e., fitted with their combined sample of SMBH and stellar-mass BH systems;
$\xi_X=0.67\pm0.12$, $\xi_M=0.78\pm0.27$, and $C=4.80\pm0.24$)
FP relations of \cite{2009ApJ...706..404G};
they are illustrated in Figure~\ref{fp_large_range}, from left to right, respectively.
Note that  in the figure the systematic offset of our measured $L_R$ values
with respect to the fiducial ones ($-11$\%; see \S2.1.1) is corrected,
and the error bars of our data points are calculated
with the error terms listed in the above (\S3.1)
in terms of the standard error propagation formula.
The low-mass AGNs match best the universal FP of \cite{2009ApJ...706..404G};
this confirms the conclusion of \citet{2014ApJ...788L..22G}.

We further test the difference between the candidate sources
(our 11 objects with 3 $<$ S/N $<$ 4.43, called Group 1)
and the well-measured sources (our 8 new objects with S/N $>$ 4.43 plus
the 9 sources from the literature excluding NGC 4395, called Group 2 here), in terms of the FP.
We calculate the orthogonal distances of every sources to the line
of the edge-on viewed universal FP of \cite{2009ApJ...706..404G}
as depicted in Figure~\ref{fp_large_range} (right panel).
The mean and standard deviation of the distances are 0.20 and 0.35, respectively, for Group 1;
0.29 and 0.26, respectively, for Group 2.
The standard errors for the two mean values are therefore 0.10 (Group 1) and 0.06 (Group 2).
Thus the difference (namely 0.09) between the mean values of the two groups is well within 1-$\sigma$ error.
We also perform Kolmogorov-Smirnov test
for the two distributions of the distances.
The resultant $p$-value (chance probability) is 0.70,
meaning that we cannot reject the hypothesis that the distributions of the two groups are the same.

\subsection{AGN radio loudness and its dependence on \mbh}

It is now generally believed that,
in non-blazar AGNs of either low or high accretion rates,
the radio emission comes predominantly from the jet,
while the high-frequency emission (from the optical through X-ray)
comes from the accretion flow
and thus is treated as an indicator of accretion rate,
namely the Eddington ratio (\lratio) in practice~\citep[e.g.][]{2003MNRAS.343L..59H,2015MNRAS.450.2317S}.
For AGNs, we can simply assume \lratio\ \mbox{$\propto L_X/ M_{\rm BH}$}.
In the literature the bolometric correction $\kappa_{\rm x}$ (defined as \lbol/$L_X$) values for AGNs
once differed considerably, and depended on \lbol\ and \lratio.
This mainly arose from the spectral complexity associated with absorption (see  Vasudevan \etal\ 2010 and references therein).
The recent studies, with various improvements in calculating the intrinsic X-ray luminosity
and particularly the bolometric luminosity,
indicate that
$\kappa_{\rm x}$ is typically in the range 10--30 derived from the observational data
with an intrinsic scatter of \mbox{$\sim$0.2\,dex},
not as large as previous deemed,
and that its dependence on either \lbol\ or \lratio\ is mild
for the observed \lratio\ regime ($\approx$10$^{-3}$ to 1);
see, e.g., Vasudevan \etal\ (2010), Brightman \etal\ (2017) and references therein.
Such a magnitude of $\kappa_{\rm x}$ variation  does not impact our  deduction here.
Thus it is easy to understand the dependence of $\mathcal{R}$ on \mbh\ and \lratio\ in terms of the FP (Eq.~\ref{FP}),
 as follows
\begin{equation}
\log\mathcal{R} = (\xi_X -1) \log {\lratio} + (\xi_M +\xi_X -1) \log M_{\rm BH}  + const. ~~ .
\label{loudness_FP}
\end{equation}
With the coefficients and their uncertainties of the universal FP of \cite{2009ApJ...706..404G},
Eq.~\ref{loudness_FP} reads:
\begin{equation}
\log\mathcal{R}  = (-0.33\pm0.12) \log {\lratio} + (0.45\pm0.30) \log M_{\rm BH}  + const.  ~~ .
\label{loudness_FP_G09}
\end{equation}
It attracts us to speculate that the radio loudness of AGNs depends not only on \lratio\
but probably also on \mbh,
even possibly to an almost equal degree
(tentatively judging from the similar magnitudes of the best-fit  power-law indexes).
The correlation might be positive with \mbh\  albeit the statistical significance being only 1.5-$\sigma$
($\mathcal{R}  \propto M_{\rm BH}^{0.45\pm0.30}$),
and negative with \lratio\  albeit the statistical significance being 3-$\sigma$
($\mathcal{R} \propto \lratio^{-0.33\pm0.12}$).
If we adopt the fitting result of \citet{2003MNRAS.345.1057M},
where the best-fit $\xi_X$, its uncertainty, and the $\xi_M$ value
are all similar to the universal relation of
\cite{2009ApJ...706..404G} but the uncertainty to $\xi_M$ is reduced by a half,
then the \lratio\ dependence would be of 3.6-$\sigma$ significance, and the
\mbh\ dependence would be of 2.4-$\sigma$ significance.
The currently large error bars on the indexes of the above FP relations
allow a considerable chance probability,
7\% (namely single-sided 1.5-$\sigma$ Gaussian deviance),
for no correlation or a negative correlation between $\mathcal{R}$ and \mbh,
thus the above speculation is yet to be verified.
On the other hand, this speculation is consistent with---and somehow reinforced by---almost all
the significant $\mathcal{R}$-related correlations in AGNs with either \mbh\ or \lratio\
that were discovered mainly by bivariate correlation analysis before
\citep[e.g.][]{2000ApJ...543L.111L,2006ApJ...636...56G}.
Furthermore, now Eq.~(\ref{loudness_FP_G09}) seems to evoke a
panoramic---and probably more insightful (see below)---understanding.

The negative $\mathcal{R}$--$\lratio$ correlation can be easily understood
under the widely accepted coupled accretion--jet models \citep{2005ApJ...629..408Y, 2003MNRAS.343L..59H},
where the accretion flow responsible for the X-ray is a hot component,
either a hot accretion flow \citep{2014ARA&A..52..529Y}
or a corona located above the cold accretion disk (\citealt{1973A&A....24..337S}).
In this model, the key factor to produce such a negative correlation is the mass accretion rate $\dot{M}$.
Hot accretion flow predicts $L_X\propto \dot{M}_{\rm BH}^{\approx2-3}$ \citep{2003MNRAS.345.1057M, 2014ARA&A..52..529Y},
while the scale-invariant jet model predicts $L_R\propto \dot{M}_{\rm BH}^{\approx1.4}$ \citep{2003MNRAS.343L..59H}.

Regarding the potentially strong and positive correlation
between $\mathcal{R}$ and \mbh, on the other hand,
it is not so easy to understand from a theoretical perspective.
Despite the currently large error bars on that index, which could be
consistent with no correlation or a negative correlation
by a chance probability of 7\% as described in the above,
we try to give an explanation for a strong, positive correlation as follows.
As \citet{2003MNRAS.343L..59H} argued,
the dependence on $\mbh$ is mainly determined by jet physics itself.
A larger $\mbh$ leads to a relatively stronger magnetic field near the BH,
which would make it easier to launch a jet.
Arguably, a stronger magnetic field strength would result in a higher acceleration,
and consequently a larger jet velocity.
This appears true observationally; i.e.,
there is likely a positive correlation between the Lorentz factor of AGN jets
$\Gamma_{\rm jet}$ and $\mbh$, as follows.
In AGNs with SMBHs, the jets are usually relativistic,
with $\Gamma_{\rm jet}\sim 10$ \citep{2004ApJ...609..539K},
whereas
in NLS1s (where \mbh\ is not too higher than the BHs in low-mass AGNs),
the jets are only mildly relativistic \citep{2015ApJS..221....3G}.
Certainly, it remains unclear whether a large-scale magnetic field
can be developed around a cold AD or not,
and thus further efforts are required,
not the least of which include better constraining any potential \mbh\  dependence of radio loudness.

\section{Summary and future work}

In studies on the fundamental plane of BH activity,
BHs in low-mass AGNs---just like the so-called intermediate-mass BHs
($\mbh \approx 10^{2-5}$\msun)---are important,
as they bridge the $\mbh$ gap between BH X-ray binaries (BHXBs) and common supermassive AGNs,
and can help to constrain the dependence on $\mbh$.
In this work, we use a new method to acquire radio flux directly from the images of the VLA FIRST survey,
for the low-mass AGNs of \cite{2012ApJ...755..167D} that have virial \mbh\ estimated from the broad \ha\ lines.
As a result, we increase the number of the low-mass AGNs with both \mbh\ estimation, firm X-ray measurement,
and radio detection of high statistical significance:
from 10 in the literature to 18,
with a total of 29 including sources with less well-constrained radio detections (see Tables 1 and 2).
Of the 19 new sources (or candidates)
out of the parent sample of 288 objects covered by the FIRST survey,
4 sources have S/N $> 5$ in radio flux
and 4 additional have S/N $>4.43$;
these 8 sources can be regarded as reliably radio-detected,
with a trial penalty (chance probability) less than  0.0014 (see \S2.1).
The other 11 sources with radio $3<$ S/N $<4.43$,
conservatively speaking, are only candidates;
in other words, one merit of this work is the target selection for future deeper radio observations.
Given the current data, we can only state that the distributions in the radio/X-ray correlation and in the FP
of the candidate radio sources are not different from
the corresponding ones of our reliable radio sources and the literature sources.

We find that most (if not all) of the low-mass AGNs follow a standard radio/X-ray correlation
(see Eq.~3 and Figure~1)
as given by \citet{2013MNRAS.428.2500C}, suggesting that they are not ``outliers".
The correlation slope between $L_R$ and $L_X$ supports the picture that
the accretion and jet processes
are quite similar in accreting systems of different BH masses.
Further, the low-mass AGNs obey the universal FP relation
fitted with the combined dataset of BHXBs and AGNs by \citet{2009ApJ...706..404G}.
In view of the FP, BH mass seems to play an important role
in determining the power of jets with respect to the accretion power;
i.e., at a given X-ray Eddington ratio ($L_X/\ledd$), systems with higher $\mbh$
tend to be
systematically brighter in radio (namely larger radio loudness $\mathcal{R}$).
If it is the case, this implies that the accretion--jet physics is mass-dependent.
In other words,  for the observed correlations concerning radio emission (see \S1),
we speculate   
a coherent picture that 
the $\mathcal{R}$ of AGNs depends not only on Eddington ratio,
but probably also on \mbh\
(even possibly to an almost equal degree). 
Certainly, this speculation is yet to be verified observationally,
since the currently large error bars on the FP indexes allow
a considerable chance probability (7\%).
Theoretically, the $M_{\rm BH}$ dependence may be related to magnetic field strength
which gets stronger with increasing $M_{\rm BH}$.

There are several lines of work for the future.
First of all,  we are proposing synthesis imaging observations
of a higher spatial resolution and a deeper depth
to pin down the exact radio emission from the nuclei
of the low-mass AGNs used in this work.
With the better data, we will be able to better constrain the FP relationship or alike of BH activity.
As a by-product, with the better-constrained FP relation we can make it clear for sure
whether the radio loudness $\mathcal{R}$ of AGNs depends on both
BH mass and Eddington ratio.
In the line of our new method to harness the VLA FIRST images,
it would be interesting to apply it to the whole data set
of low-$z$ Seyfert galaxies in the SDSS,
trying to address why the jets in Seyfert galaxies cannot be fully developed.
Meanwhile, we will find more radio-detected low-mass AGNs,
enabling the update of the present work.

\acknowledgments
We thank the anonymous referee for the thorough and helpful comments and suggestions
(including improving the English presentation),
particularly for his/her analyzing our S/N criterion from the perspective of trial penalty.
We thank Prof. Qizhou Zhang and Qian Long for helpful discussions.
This work is supported by National Key R\&D Program of China No. 2017YFA0402600,
State Key Development Program for Basic Research
(2015CB857100),
Natural Science Foundation of China grants (NSFC No.~11473062 
and 11603036), and the Open Project Program of the Key Laboratory of FAST, NAOC, Chinese Academy of Sciences.
FGX is supported in part by the National Key Research and Development Program of China (2016YFA0400804),
the Youth Innovation Promotion Association of CAS (id.~2016243), and the Natural Science Foundation of Shanghai (No.~17ZR1435800).
This research has made use of the NASA/IPAC Extragalactic Database (NED) which is operated by
the Jet Propulsion Laboratory, California Institute of Technology,
under contract with the National Aeronautics and Space Administration.



\clearpage

\clearpage

\begin{table*}[htbp]
\caption{Properties of our measured low-mass AGNs for FP research}
\tiny
\label{results}
\begin{center}
\begin{tabular}{|c|c|c|c|c|c|c|c|c|}
\hline
Name & redshift & S/N$_{\rm 1.4GHz}$ & $f_{\rm 1.4 GHz} $ & $\log \nu L_{\nu}(5\ \rm GHz) $  & $f_{\rm 2-10 keV}$ & $\log L_{\rm 2-10 keV}$ & $\log M_{\rm BH}$ & Remark\\
        &              &                                  &      (mJy)                   &   (erg s$^{-1}$)                 &   (erg s$^{-1}$ cm$^{2}$)   &   (erg s$^{-1}$)             &      &       \\
(1)    & (2)        &   (3)                           &           (4)                  &                     (5)                 &                                   (6)    &                    (7)             & (8) & (9)  \\
\hline
J000111.15-100155.7 & 0.0489 & 3.0  & 0.49$\pm$0.08  & 37.87$^{+0.08}_{-0.11}$ & $(1.05\pm 0.23)\times 10^{-13}$ & 41.80$^{+0.10}_{-0.13}$ & 6.2 & 1 \\
 J030417.78+002827.4 & 0.0445 & 3.4 & 0.62$\pm$0.11  & 37.88$^{+0.09}_{-0.11}$ & $(3.66\pm 0.14)\times 10^{-12}$ & 43.26$^{+0.06}_{-0.07}$ & 6.2 & 1 \\
 J032606.77+011429.7 & 0.1274 & 4.8 & 0.63$\pm$0.08  & 38.80$^{+0.07}_{-0.09}$ & $(3.52\pm 1.49)\times 10^{-13}$ & 43.16$^{+0.16}_{-0.26}$ & 6.3 & 3\\ 
 J073106.87+392644.7 & 0.0485 & 4.5 & 0.61$\pm$0.08  & 37.95$^{+0.08}_{-0.09}$ & $(2.38\pm 0.12)\times 10^{-12}$ & 43.15$^{+0.06}_{-0.07}$ & 6.0 & 1 \\
 J081550.24+250641.0 & 0.0727 & 3.8 & 0.59$\pm$0.09  & 38.29$^{+0.08}_{-0.10}$ & $(2.27\pm 0.35)\times 10^{-13}$ & 42.48$^{+0.08}_{-0.11}$ & 5.9 & 1 \\
 J082433.33+380013.1 & 0.1031 & 5.5 & 1.07$\pm$0.15  & 38.84$^{+0.08}_{-0.10}$ & $(9.43\pm 0.45)\times 10^{-14}$ & 42.40$^{+0.06}_{-0.07}$ & 6.1 & 2, 4 \\
 J085152.63+522833.0 & 0.0645 & 4.8 & 0.92$\pm$0.14  & 38.37$^{+0.08}_{-0.10}$ & $(3.95\pm 0.47)\times 10^{-13}$ & 42.62$^{+0.08}_{-0.09}$ & 5.8 & 1 \\
 J104504.24+114508.8 & 0.0548 & 4.1 & 0.82$\pm$0.14  & 38.18$^{+0.09}_{-0.11}$ & $(1.60\pm 0.09)\times 10^{-13}$ & 42.08$^{+0.06}_{-0.08}$ & 6.2 & 1 \\
 J105131.91+504223.2 & 0.1321 & 3.6 & 0.54$\pm$0.08  & 38.77$^{+0.08}_{-0.10}$ & $(2.71\pm 0.74)\times 10^{-13}$ & 43.08$^{+0.12}_{-0.16}$ & 6.1 & 3\\
J110258.74+463811.5 & 0.1490 & 3.1 & 0.75$\pm$0.16  & 39.01$^{+0.10}_{-0.13}$ & $(2.84\pm 0.75)\times 10^{-13}$ & 43.20$^{+0.12}_{-0.16}$ & 6.0 & 3\\
 J131926.53+105611.0 & 0.0643 & 3.5 & 0.78$\pm$0.15 & 38.30$^{+0.10}_{-0.12}$ & $(3.73\pm 1.43)\times 10^{-14}$ & 41.60$^{+0.15}_{-0.23}$ & 5.9 & 1 \\
J133928.50+403229.9 & 0.1179 & 3.2 & 0.66$\pm$0.13  & 38.76$^{+0.10}_{-0.12}$ & $(5.45\pm 0.47)\times 10^{-12}$ & 44.28$^{+0.07}_{-0.08}$ & 6.0 & 3\\
 J134249.27+482723.7 & 0.0912 & 3.3 & 0.68$\pm$0.13  & 38.55$^{+0.09}_{-0.12}$ & $(1.62\pm 0.55)\times 10^{-13}$ & 42.53$^{+0.14}_{-0.20}$ & 6.1 & 3\\
 J134738.24+474301.9 & 0.0641 & 5.4 & 0.84$\pm$0.11  & 38.33$^{+0.08}_{-0.09}$ & $(8.91\pm 0.71)\times 10^{-13}$ & 42.97$^{+0.07}_{-0.08}$ & 5.6 & 1, 4 \\
 J140040.57-015518.3 & 0.0250 & 9.4 & 1.75$\pm$0.16  & 37.82$^{+0.07}_{-0.08}$ & $(4.42\pm 0.49)\times 10^{-14}$ & 40.84$^{+0.07}_{-0.09}$ & 6.3 & 2\\
 J141234.67-003500.1 & 0.1270 & 5.8 & 0.54$\pm$0.05  & 38.73$^{+0.07}_{-0.08}$ & $(5.84\pm 0.53)\times 10^{-13}$ & 43.38$^{+0.07}_{-0.08}$ & 5.9 & 1 \\
 J143310.55+525830.5 & 0.0474 & 4.8 & 0.83$\pm$0.12  & 38.06$^{+0.07}_{-0.10}$ & $(6.78\pm 1.97)\times 10^{-14}$ & 41.59$^{+0.12}_{-0.17}$ & 6.0 & 1 \\
 J144108.70+351958.8 & 0.0792 & 3.4 & 0.72$\pm$0.14  & 38.45$^{+0.08}_{-0.12}$ & $(1.17\pm 0.18)\times 10^{-13}$ & 42.27$^{+0.08}_{-0.11}$ & 5.8 & 2\\
 J150752.53+515111.1 & 0.0748 & 4.2 & 0.63$\pm$0.09  & 38.34$^{+0.10}_{-0.10}$ & $(7.04\pm 0.78)\times 10^{-13}$ & 43.00$^{+0.07}_{-0.09}$ & 6.2 & 3\\
\hline
\end{tabular}
\end{center}
\scriptsize
Note. -- Col. (1): Official SDSS name in J2000.0.
Col. (2): Redshift measured by the SDSS pipeline.
Col. (3): S/N (namely flux/rms) of our 1.4\,GHz measurement; here the flux is not corrected for the CLEAN bias (see \S2.1.1).
Col. (4): Our measured 1.4\,GHz flux corrected for the CLEAN bias, and the rms noise.
Col. (5): Radio luminosity and its 1-$\sigma$ error. The error includes the rms noise,
a 13\% error from the uncertainty of $\alpha_R$, and a 10\% statistical error of our measured radio fluxes with respect to the fiducial ones (see the end of \S2.1.1).
Col. (6): The 2--10 keV flux, and the 1-$\sigma$ error calculated from the HEASARC uncertainty of flux or count rate.
Col. (7): X-ray luminosity and its 1-$\sigma$ error. The error includes the observational uncertainty (listed in Col.~6)
and a 15\% error from the uncertainty in the photon index.
Col. (8): Black hole mass as listed in Dong \etal\ (2012; their Table~3).
Col. (9): Remarks: 1. X-ray flux from Chandra.
2. X-ray flux from XMM-Newton.
3. X-ray flux from ROSAT.
4. Also included in \cite{2015ApJ...808..163P}.
\end{table*}

\begin{table*}[htbp]
\caption{Properties of the literature low-mass AGNs for FP research}
\tiny
\label{literaturetable}
\begin{center}
\begin{tabular}{|c|c|c|c|c|c|c|c|}
\hline
Name & $D$    & $\log \nu L_{\nu}(5\ \rm GHz) $ &  $\log L_{\rm 2-10 keV}$ & $\log M_{\rm BH}$ & Reference\\
          &  (Mpc) & (erg s$^{-1}$) &    (erg s$^{-1}$) &       &    \\
(1)     &   (2)     &   (3)                &           (4)            & (5) &  (6)    \\
\hline
J082443.28+295923.5 &  115.0 & 38.03$\pm$0.02 &  42.51$^{+0.08}_{-0.04}$  &  5.70  & 1 \\
J091449.05+085321.1 &  644.4 & 39.33$\pm$0.02 &  43.26$\pm$0.03  &  6.30  & 1 \\
J101246.49+061604.7 &  359.0 & 38.67$\pm$0.03 &  42.56$\pm$0.04  &  6.22  & 1 \\
J110501.97+594103.6 &  156.5 & 38.37$\pm$0.02 &  42.30$^{+0.05}_{-0.09}$  &  5.58  & 1 \\
J124035.81-002919.4 (GH10) &  372.8 & 38.73$\pm$0.03 &  42.28$^{+0.14}_{-0.09}$  &  6.35  & 1, 3 \\
J132428.24+044629.6 &   96.7 & 37.61$\pm$0.03 &  41.25$\pm$0.07  &  5.81  & 1 \\
J155909.62+350147.4 &  142.7 & 38.01$\pm$0.03 &  42.77$\pm$0.03  &  6.31  & 1 \\
NGC 404 & 3.1 & 34.58$\pm$0.04 &   37.10$^{+0.20}_{-0.18}$ &  5.65 & 2, 3 \\
NGC 4395 & 4.3 & 34.95$\pm$0.06 &  39.91$^{+0.09}_{-0.12}$ &  5.56 & 3, 4, 5 \\
Henize 2-10 & 9.0 & 35.61$\pm$0.05 & 38.11$^{+0.05}_{-0.16}$ & 5.90 & 3, 6, 7 \\
\hline
\end{tabular}
\end{center}
\scriptsize
Note. -- Col. (1): Source name.
Col. (2): Distance.
Col. (3): Radio luminosity $\nu L_\nu$(5\,GHz).
Col. (4): X-ray luminosity $L_X$(2--10\,keV).
Col. (5): Black hole mass.
Col. (6): Reference:
1. \cite{2014ApJ...788L..22G}; 2. \cite{2010ApJ...714..713S};
3. \cite{2012ApJ...753..103N}; 4. \cite{2001ApJS..133...77H};
5. \cite{2005AJ....129.2108M}; 6. \cite{2016ApJ...830L..35R}; 7. \cite{2012ApJ...750L..24R}.
\end{table*}

\begin{table}[htbp]
\caption{Comparison of Radio Flux}
\tiny
\label{radio_flux}
\begin{center}
\begin{tabular}{|c|c|c|}
\hline
Name & $f_{\rm 1.4 GHz} $(our) & $f_{\rm 1.4 GHz}$(cat) \\
          &    (mJy)      &           (mJy)       \\
(1)     &   (2)            &           (3)           \\
\hline
J121629.13+601823.5 & 0.94$\pm$0.13 & 1.03$\pm$0.11  \\
J162824.49+452811.0 & 0.75$\pm$0.09 & 0.68$\pm$0.09  \\
J074251.09+333403.9 & 2.75$\pm$0.14 & 3.18$\pm$0.16  \\
J074948.33+264734.2 & 3.55$\pm$0.09 & 4.53$\pm$0.13  \\
J132428.24+044629.7 & 1.79$\pm$0.13 & 2.33$\pm$0.14  \\
J132834.37-030744.8 & 7.10$\pm$0.13 & 8.81$\pm$0.14  \\
\hline
J082443.28+295923.5 & 1.67$\pm$0.16 & $1.77\pm 0.11$   \\
J101246.49+061604.7 & 1.04$\pm$0.11 & $1.03\pm 0.15$   \\
J121629.13+601823.5 & 0.94$\pm$0.13 & $1.03\pm 0.11$   \\
J132428.24+044629.6 & 1.79$\pm$0.15 & $2.33\pm 0.14$   \\
J155909.62+350147.4 & 2.72$\pm$0.11 & $3.39\pm 0.12$   \\
NGC  4395 & 1.12$\pm$0.12 & $1.17\pm 0.15$ \\
\hline
\end{tabular}
\end{center}
\scriptsize
Note. -- Col. (1): Source name.
Col. (2): Our measured 1.4\,GHz flux (corrected for the CLEAN bias) and rms noise.
Col. (3): The 1.4\,GHz flux and the rms retrieved from the FIRST catalog.
The upper part lists the 6 unresolved sources among the 17 low-mass AGNs included in the FIRST catalog,
while the lower part lists the 6 sources that are in the literature dataset (Table~2)
and covered by the FIRST survey (but excluding  those with  \textit{Deconv.\,MajAx} $\geqslant 1\farcs5$ in the FIRST catalog);
see the text in \S2.1.1 for the detail.
\end{table}

\end{document}